\documentclass[twocolumn]{aastex62}

\newcommand{\FWHM}{\textnormal{FWHM}}

\graphicspath{{./}{figures/}}

\shorttitle{Modeling High-Cadence FRB Surveys}
\shortauthors{McGregor and Lorimer}

\begin{document}

\title{Modeling Current and Future High-Cadence Surveys of Repeating FRB Populations}

\author[0000-0003-2111-3437]{Kyle McGregor}
\affil{Wesleyan University Department of Astronomy and Van Vleck Observatory, 96 Foss Hill Dr., Middletown CT 06459, USA}
\affil{Department of Physics and Astronomy, West Virginia University, P.O. Box 6315, Morgantown, WV 26506, USA}

\author[0000-0003-1301-966X]{Duncan R. Lorimer}
\affil{Department of Physics and Astronomy, West Virginia University, P.O. Box 6315, Morgantown, WV 26506, USA}
\affil{Center for Gravitational Waves and Cosmology, West Virginia University, Chestnut Ridge Research Building, Morgantown, WV 26505, USA}



\begin{abstract}

In recent years, the CHIME (Canadian Hydrogen Intensity Mapping Experiment) interferometer has revealed a large number of Fast Radio Bursts (FRBs), including a sizable population that demonstrates repeating behavior. This transit facility, employing a real-time FRB search pipeline, continually scans the sky with declinations between $-10^{\circ}$ and $90^{\circ}$ for events with fluences $\gtrapprox 0.4$~Jy~ms. We simulate a population of repeating FRBs  by performing Monte Carlo simulations of underlying source populations processed through a mock CHIME/FRB observing pipeline. Assuming intrinsic repeater rates follow a Poisson distribution, we  test assumptions about the  burst populations of the repeater sample, and construct models of the FRB sample assuming various cosmological distributions.  We infer the completeness of CHIME/FRB observations as a function of observing cadence and redshifts out to 0.5. We find that, if all simulated bursts have a fixed Poisson probability of repetition over their integrated time of observation, repeating burst detections across comoving volume should continue to grow near linearly on the order of decades. We predict that around 170 of the current CHIME/FRB one-off sources will ultimately repeat. We also make projections for FRB repeaters by future facilities and demonstrate that the number of repeaters they find could saturate on a $\sim 3$~yr timescale.

\end{abstract}



\section{Introduction} \label{sec:intro}
Fast Radio Bursts (FRBs) are among the most enigmatic extragalactic sources of radiation. First discovered in 2007 using archival data from the Parkes (Murriyang) telescope  \citep{LorimerBurst} these dispersed highly-energetic millisecond-scale bursts of radio emission have been shown to be consistent with a cosmological origin \citep{2013Sci...341...53T}. While no conclusive progenitor mechanism for FRB emission has been widely accepted,  modeling shows that directional magnetar flares  have energy budgets similar to observed burst energies \citep[see, e.g.,][]{2018PhyU...61..965P}. 

 With 672 one-off bursts currently known\footnote{For a list, see https://blinkverse.alkaidos.cn.} \citep{Xu_2023}, a recent data release from the Canadian Hydrogen Intensity Mapping Experiment (CHIME) collaboration has more than doubled the number of known repeating FRBs, currently standing at 63 \citep{rn3}. Searches led by the CHIME/FRB collaboration using this facility are carrying out a census of FRBs. As a transit instrument that each day observes the 200 square degrees flanking the north-south meridian of the sky at declinations between --10 and 90 degrees, CHIME repeatedly scans the sky and is therefore well suited to probe the repeating FRB population.

With a small known sample size compared to the inferred daily incidence of bursts from across the universe \citep[expected to be on the order of thousands per day;][]{2013Sci...341...53T}, a Monte Carlo simulation is a powerful tool to infer properties of the underlying FRB population(s). It is also important to determine the limits of our facilities and find how many bursts can be found in a given cadence, as well as account for survey selection effects that may influence completeness. We present a model for an observed repeater population constructed from first principles, which can provide a prediction for future detection rates by CHIME/FRB. From this, we can determine a model census of bursts whose properties can be compared to the known sample revealed in \citep{rn3}.

Inspired by the latest release of the repeater sample \citep{rn3}, we apply a Monte Carlo approach in this paper. In Section \ref{build} we discuss how we build a model source population, in Section \ref{observe} we discuss our procedure for generating a mock CHIME/FRB observing campaign on this population. We compare these modeled populations to the observed population in Section \ref{results}, and discuss the relevance to previous and future work in Section \ref{discussion} and present our conclusions in Section \ref{conclusion}.

\section{Model Source Properties}
\label{build}

In constructing a synthetic population of repeaters, we must first consider properties intrinsic to each source. These parameters include right ascension and declination on the sky ($\alpha$, $\delta$), redshift  ($z$), mean luminosity ($L_0$), intrinsic pulse width ($w_{\rm int}$), and host-galaxy dispersion measure ($\rm{DM}_{Host}$). 

As FRBs are a cosmological population, we assume a uniform distribution of locations on the celestial sphere. We draw from a uniform random deviate between 0--24~hr in right ascension. In declination space, however, following \citet{ChawlaPaper}, we draw uniformly across $-10^{\circ} < \delta < 90^{\circ}$. The cosine dependence in solid angle for an isotropic population cancels the inverse cosine dependence in exposure. While not representing a uniform distribution on the celestial sphere, this does give a distribution proportional to the product of the declination-dependent sky coverage and exposure for CHIME/FRB which more accurately represents the underlying position distribution of detected sources \citep[for details, see][]{ChawlaPaper}.

It is currently unclear what the underlying redshift distribution of FRBs is in general, and there is debate as to the distribution probed by CHIME/FRB \citep[see, e.g.,][]{2022ApJ...924L..14Z,2023ApJ...944..105S}. In addition, the DM distributions for CHIME/FRB repeaters appears to be significantly different to those of the one-off bursts \citep{rn3}, which leads to further uncertainty in their underlying redshift distributions. For our models of repeating FRBs,
we explore two scenarios for the redshift distributions: a population following a constant density in comoving volume ($V_c$) and a population following the cosmic star formation rate \citep[SFR;][]{MadauPaper}. For the former case, the probability density in redshift
\begin{equation}
   P_{\rm comoving}(z) \propto  \left( \frac{1}{1+z}\right) \frac{dV_c}{d\Omega dz},
\label{ConstComoving}
\end{equation}
where  $\Omega$ is the solid angle and $\frac{dV_c}{d\Omega dz}$ is the differential comoving volume element. For the latter case, 
\begin{equation}
    P_{\rm SFR} (z) \propto  \left( \frac{(1+z)^{2.7}}{1+\left[\frac{(1+z)}{2.9}\right]^{5.6}} \right)  \,
    \left(\frac{1}{1+z}\right) \,
    \frac{dV_c}{d\Omega dz} \cdot 
\label{cSFR}
\end{equation}
In both cases, as we discuss later, we restrict our maximum redshifts probed by the simulations to be $z_{\rm max}=0.5$. We initially created simulations with FRBs out to higher redshift limits, but found that too many high-DM sources were produced as a result. Our choice of $z_{\rm max}=0.5$ provided the best match of the DMs seen in the CHIME/FRB catalogs to date \citep{cat1,rn3}.

For the mean luminosity of each burst, $L_0$, we draw from a Schechter function employing a fit from \citet{2023ApJ...944..105S}.  We impose bounds between $10^{39}$ and $10^{45}$~erg~s$^{-1}$. For a 1~ms pulse, this is in line with the inferred luminosity limits of observed bursts, consistent with the burst energy limits used by \citet{ChawlaThesis}. The Schechter luminosity distribution is commonly used to describe the energetics of extragalactic sources and takes the form
\begin{equation}
    P_{L_0}(L_0)dL_0 = \left(\frac{L_0}{L_*}\right)^{\alpha+1}\exp{\left(\frac{L_0}{L_*}\right)} d(\log L_0),
\end{equation}
where the characteristic luminosity $L^* = 2 \times 10^{44}$~erg~s$^{-1}$ and power law slope $\alpha = -1.3$. Given that the current CHIME/FRB repeaters are known to vary in intensity \citep{rn3}, we dither the daily burst luminosities using a scheme described in Section 3.

In simulating intrinsic pulse widths, we draw from a lognormal distribution as given in \citet{PulseWidthLuminosityII}. The dispersion measure (DM) of a burst is the integral of electron number density over the line of sight to the burst, which will include contributions from each source of propagation en route to Earth and thus serves as an observable proxy for distance. For our simulations, following previous authors \cite[e.g.,][]{LuminosityFunction}, we computed the observed DM,
\begin{equation}
\label{eqn:DM}
\rm
DM_{obs} = DM_{MW}(\textit{l},\textit{b})+DM_{IGM}(\textit{z})+\frac{DM_{host}}{1+\textit{z}},
\end{equation}
with the Galactic component, DM$_{\rm MW}$, from the YMW16 model of Milky Way electron density \citep{YMW16}, the intergalactic medium (IGM) component, DM$_{\rm IGM}$, from the Macquart ($z$-DM) relation \citep{MacquartRelation}, and the repeater host galaxy contribution, DM$_{\rm host}$, from a lognormal random deviate parametrized by \citet{HostGalaxyDispersion}.

We draw a catalog of $10^5$ mock repeaters using Monte Carlo methods in \texttt{scipy}. This gives us a base of source repeater progenitors over which we can iterate to simulate an observation campaign with CHIME, the procedure for which is described in the next section.

\section{Model CHIME Observing Campaign}
\label{observe}

While CHIME is a complex instrument, responsive but highly sensitive to on-site conditions that are prohibitive to model efficiently \citep[see, e.g.,][]{FluxCalibrationCHIME}, we can infer the observability of a given event by imposing a fluence threshold of 0.4 Jy ms. This lower limit is consistent across the CHIME Catalog 1 data \citep{cat1} and the new repeater data \citep{rn3}, and used in similar population syntheses \citep{ChawlaThesis,Yamasaki+23}. 

\begin{figure}[ht]
    \centering
    \includegraphics[width=0.5\textwidth]{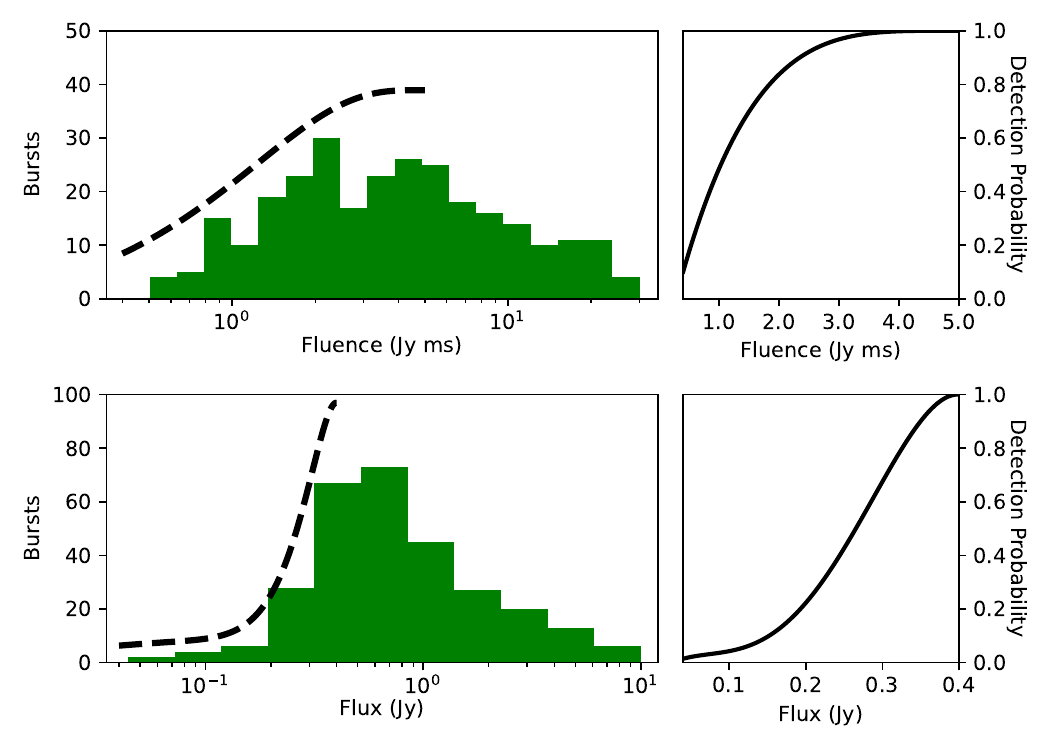}
    \caption{Fluence (top) and flux (bottom) distributions for the CHIME/FRB repeater sample (green histograms) showing the drop in the number of bursts around the threshold. These were chosen to be 5~Jy~ms in fluence and 0.4~Jy in peak flux, respectively (see text).
    Dashed and solid lines show polynomial fits to these tails in log and linear spaces, respectively. Normalization of these fits on the right to a probability scale allows us to implement these sensitivity roll offs in our simulated FRBs (see text).}
\end{figure}

\begin{figure*}
    \centering
    \includegraphics[width=0.95\textwidth]{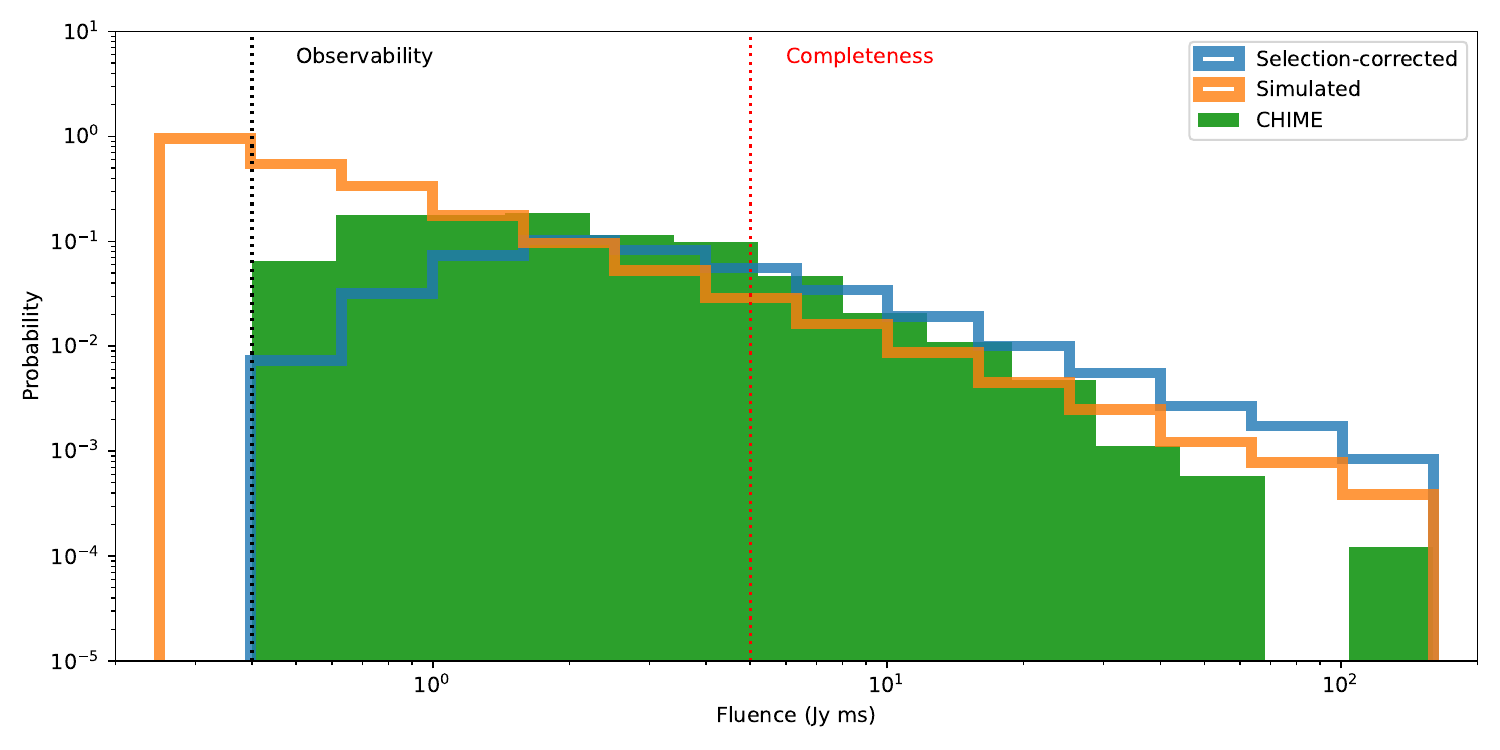}
    \caption{Observed CHIME/FRB fluence distributions compared to those from our model population. The simulated distribution  (orange) shows the constant \textbf{star} formation model with the simple 0.4~Jy~ms fluence threshold applied. The selection-corrected distribution  (blue) shows just those FRBs which survived the Bernoulli trials in flux and fluence (see text).}
\end{figure*}

Our mock observation pipeline first iterates over each of the $10^5$ simulated repeater and determines the daily burst incidence using a Poisson random deviate. Due to the shape of CHIME's primary beam, higher declinations will have a greater exposure over CHIME's operations, resulting in a more complete sample at high declinations. This declination dependence has been modeled in similar studies \citep[see][]{ClancyJames}. We choose to weigh the daily Poisson burst incidence by this exposure in our forward modeling, such that \textbf{sources at} higher declinations are more likely to \textbf{be observable} on a given day to correct for the declination bias. The mean of the probability mass function for the number of observed bursts each day over the four CHIME cylinders,
\begin{equation}
\lambda = \frac{4\times \textrm{FWHM}_{E-W}}{360^{\circ}\cos(\delta)} \,\,
\Gamma(R_{\rm min},R_{\rm max},\gamma),
\end{equation}
with $\FWHM_{\rm E-W} = 0.32^\circ$ the full width at half maximum of each of synthesized beam \citep{ChawlaThesis}. This corresponds to the product $\lambda = (\textrm{time in beam}) \times (\textrm{bursts day}^{-1})$ for each day of the simulation.
Following
\citet{ClancyJames},
$\Gamma(R_{\rm min},R_{\rm max},\gamma)$ is a power-law random deviate ranging between $R_{\rm min}=10^{-3}$ and $R_{\rm max}=10^{-1}$ $\rm bursts~day^{-1}$
with a power-law index
$\gamma=-2.2$. We choose $R_{\rm min}$ to be approximately the reciprocal of the current survey length (3.3~yr). We found that models with $R_{\rm max}>0.1$~day$^{-1}$ produce too many observable events.  These are defined in terms of the total survey cadence, rather than the time in beam. In addition, we do not assume any correlation between burst rate and intrinsic luminosity, $L_0$, over the range of $L_0$ we model.

The observed pulse width, $w_{\rm obs}$, includes contributions from sources of propagation and instrumental parameters. It is given by the quadrature addition 
\begin{equation}
    w_{\rm obs} = \sqrt{[(1+z)w_{\rm int}]^2+w_{\rm scat}^2+w_{\rm samp}^2+w_{\rm DM}^2}\,,
\end{equation}
where $w_{\rm int}$ is the intrinsic pulse width and the $(1+z)$ factor accounts for time dilation, $w_{\rm scat}$ is the scatter broadening, $w_{\rm samp}$ is the data sampling interval and $w_{\rm DM}$ is the dispersion broadening across a finite frequency channel. To model the intrinsic width, following \citet{PulseWidthLuminosityII}, we draw from a log normal distribution so that 
log$_{10} \, (w_{\rm int}/{\rm ms})=$~normal($\mu_W,\sigma_W$).
Here the normal distribution has a mean $\mu_W = 0.2$ and standard deviation $\sigma_W=0.33$.
The CHIME/FRB sampling time $w_{\rm samp} = 0.98$~ms \citep{CHIMEOverview}. 
The scattering timescale is calculated alongside the $\rm DM_{MW}$ and $\rm DM_{IGM}$ contributions with the \texttt{pygedm} package \citep{pygedm} using the electron density model developed by \citet{YMW16}. 

As mentioned above, some variation in the flux density of repeaters across observations is noted in the CHIME sample, which implies a mechanism for differing luminosity over consecutive bursts from the same repeater. While the sample size for most observed bursts is much too small to easily parametrize, we assume a normal distribution with a standard deviation $\sigma = 0.1L_0$. Using Eq.~8 of \citet{FluenceRedshiftDMs}, we find that the observed fluence
\begin{equation}
    F_\nu = \frac{(1+z) \, w_{\rm int}}{4\pi D_L^2 \Delta \nu} \,\,{\rm normal}(L_0, 0.1L_0),
\end{equation}
where $D_L$ is the luminosity distance and the bandwidth
for CHIME/FRB, $\Delta \nu = 400$~MHz. Notably here, for simplicity, we are being agnostic about any dependence of the fluence with frequency and implicitly assume that the FRBs are flat spectrum sources over the frequency ranges explored here (i.e., 400~MHz out to $(1+z_{\rm max})\,800$~MHz = 1200~MHz).

As noted by \citet{2003ApJ...596.1142C}, the signal-to-noise threshold for FRB detection can be computed through the radiometer noise considerations. Originally, we followed this approach and implemented a discretization of CHIME/FRB's beam to infer the observability of a given FRB using the radiometer equation. This method is unphysical in that it assumes a fiducial nature to CHIME's system-equivalent flux density, which in reality is calibrated in real-time as part of CHIME/FRB's \texttt{bonsai} FRB search algorithm. Following the detailed sensitivity determinations by \citet{ChawlaThesis}, however, after iterating over each day for each repeater in the population, we accept a burst as observable if its fluence is above the 0.4~Jy~ms threshold. This gives us a population theoretically observable by CHIME, though selection corrected for various incompleteness factors compared to the CHIME survey populations.

To accurately model CHIME's observing campaign, we must reintroduce these incompleteness factors which become important close to the detection threshold. It is known that CHIME is observationally biased against low-fluence events, which can only be detected at the highest gain regions of each synthesized beam, whereas higher fluence events can be readily detected across the primary beam and sidelobes \citep{DoAllBurstsRepeat?}. We account for this bias by weighting the probability of observation for each low-fluence and low-flux event with an independent Bernoulli trial, which will return a ``success'' or ``failure'' based on the exhausted probability of detection at each fluence and flux in the over-represented regime. We employ a polynomial fit for the drop-off following the empirical cutoff curve in the observed fluence/flux distributions between the 0.4 Jy ms lower fluence limit and the 5 Jy ms $95\%$ completeness threshold identified in \citet{cat1}. The flux threshold was empirically determined so that its range was approximately an order of magnitude above the minimum observed flux in the catalog.
This is done during the filtering step after the mock observed catalog is collected in order to aid computational efficiency. 

The fourth-degree polynomial fits to the left edges of the fluence and flux distributions are shown in Fig.~1. We report the normalized fits as $p(F_\nu) = -0.002(F_\nu-5)^4+1$ for fluence and $p(S_\nu) = -240.8 S_\nu^4 + 174.4S_\nu^3 - 30.85 S_\nu^2 + 2.46 S_\nu - 0.045$ for flux density where the units of fluence and flux density at Jy~ms and Jy, respectively. In Fig.~2 we show the application of this approach to the constant star formation model. In general, the better agreement between the ``Observability'' and ``Completeness'' thresholds is clearly evident. We note, however, that the CHIME/FRB sample at low fluences has a different shape than our selection-corrected distribution. We believe the reason for this is reflective of the fact that the CHIME/FRB fluences are lower limits due to their uncertain position within the telescope beam which we do not model. We discuss this point further below. At higher fluences, additionally, the discrepancy seen is reflective of the small sample size of such repeaters observed to date.

\section{Results}
\label{results}

\begin{figure*}[!th]
    \centering
    \includegraphics[width=\textwidth]{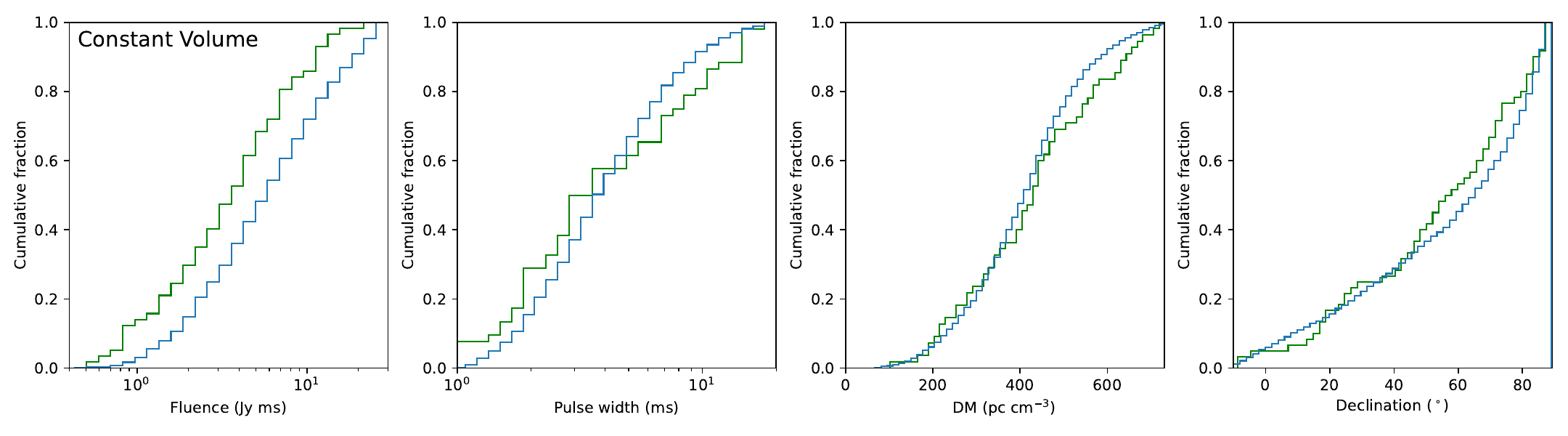}
    \includegraphics[width=\textwidth]{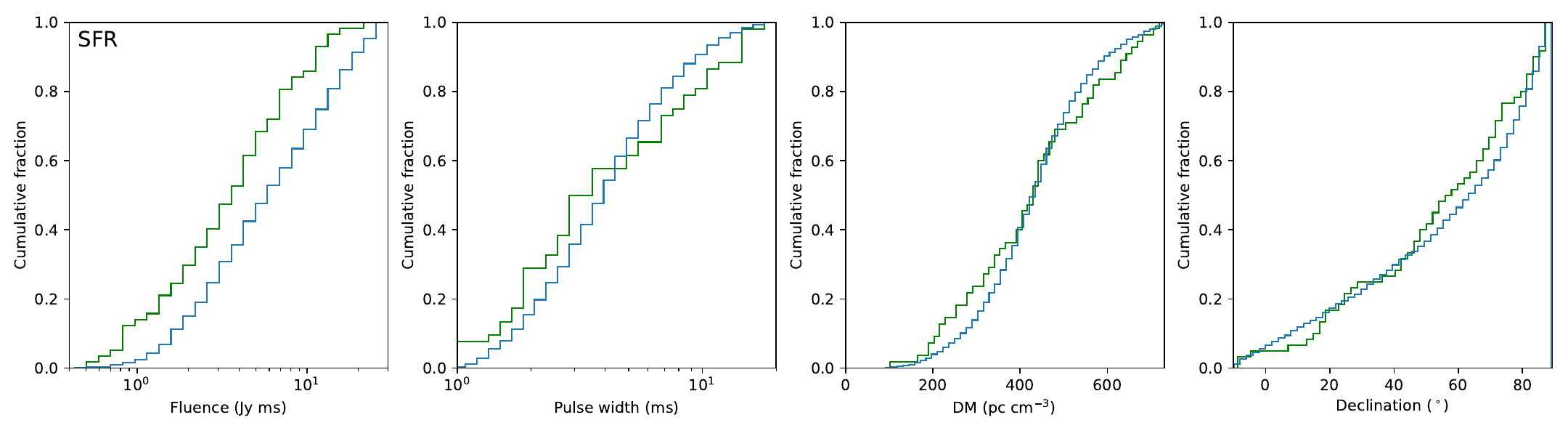}
    
    \caption{Cumulative distributions for the redshift distribution tracing the constant volume redshift distribution (Eq.~1; upper row) and the 
    cosmic star formation rate  (Eq.~2; lower row). Green lines show the observed distributions from CHIME/FRBs. The blue lines are our simulated observable populations.}
    \label{fig:CDFs}
\end{figure*}

We ran the simulations above for a total of 7300 days (20 years), extracting the sample at 1095 days (three years) to match the timescale of the CHIME sample \citep{rn3}. Fig.~\ref{fig:CDFs} show cumulative distributions for the observed and model samples
for the constant star formation and constant volume redshift distributions, respectively. For each model, we show the fluence, pulse width, DM, and declination distributions. While we do not take the approach of fitting models to the data in this paper, it is useful to quote quantitative metrics of comparison between the model and observed samples. Table 1 shows the results of the two-sided Kolmogorov-Smirnov (KS) tests \citep{kolmogorov1933sulla,smirnov1948table} we ran on these distributions.

\renewcommand{\arraystretch}{1.5}
\begin{deluxetable}{ l c c c c c }
\label{table:KStests}
\centering
\tablewidth{0pt}  
\tablecaption{KS-test results for the two redshift models considered in this study. For each model, we provide the KS statistic and associated $p$-value for fluence, observed pulse width, DM and declination, respectively.
}
\tablehead{
    \colhead{Model} & \colhead{} &\colhead{$F_\nu$} & \colhead{$w_{\rm obs}$} &  \colhead{DM}  & \colhead{Decl.}}
\startdata
    Comoving  & KS statistic  & $0.383$   & $0.189$  & $0.112$ & $0.154$ \\
                    & $p$-value   & $2\times10^{-8}$   & $0.030$  & $0.398$ & $0.112$ \\
\hline
    SFR & KS statistic  & $0.406$   & $0.185$  & $0.104$ & $0.146$ \\
                   & $p$-value   & $3\times10^{-9}$   & $0.035$  & $0.565$ & $0.147$ \\
\enddata
\end{deluxetable}

\section{Discussion}
\label{discussion}

The goal of this work is to constrain some of the properties of the FRB repeater population as seen through regular observations with the CHIME/FRB survey described by \citet{rn3}. In doing this, we have tried to strike a balance between a simple analytic model and a very detailed simulation which would also include Markov Chain Monte Carlo techniques for parameter estimation. Such a simulation procedure has been followed in concurrence with this work by \citet{Yamasaki+23}, investigating the CHIME/FRB source count evolution with a growing repeater population, yielding similar parameters. In many respects, we have taken an approach similar to earlier studies of the pulsar population \citep[see, e.g.,][]{2006ApJ...643..332F} where plausible models are developed that can mimic the main results. In the sections below, we discuss the main findings from this approach.

\subsection{Which model is better?}

As can be seen in Fig.~\ref{fig:CDFs}, and summarized quantitatively in Table 1, the agreement between our model and observed FRB populations is generally good and we briefly comment on each of the distributions considered. While the model with the redshift distribution tracing the cosmic star formation matches the DM distribution slightly better than the constant volume model, the KS tests indicate that the two models are statistically very similar in their match to the observed CHIME/FRB repeater DM distribution. This is also reflected in the other distributions and KS test scores and is generally in lines with our expectations, since the redshift distributions we considered have a similar behaviour in the range $0<z<0.5$. For the fluence distribution, we note that our model fluences are systematically higher than those observed by CHIME/FRB. We did not seek to optimize this further, given that the fluences in the CHIME/FRB catalog are lower limits due to uncertainties in the true source position within the CHIME beam.
In the remainder of the discussion, we use the cosmic star formation redshift distribution (Eq.~2) as our reference model.

\subsection{Repeater detections versus time}

\begin{figure}[ht]
    \centering
    \includegraphics[width=0.45\textwidth]{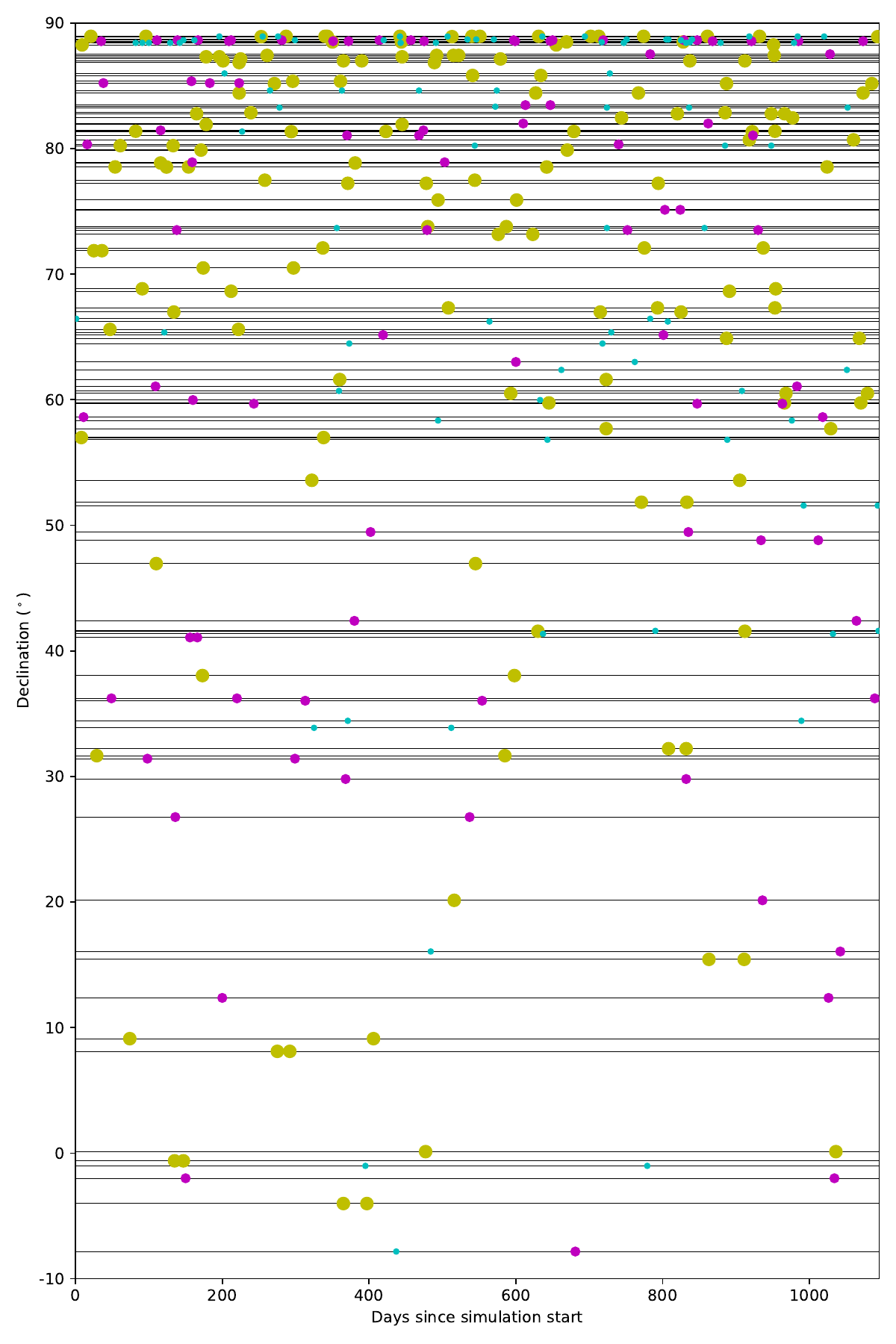}
    \caption{Declination versus time for 163 selected repeaters from the simulated population. Cyan represents bursts with $ F_\nu < 4$ Jy ms, magenta $4 \textrm{ Jy ms} < F_\nu < 40$ Jy ms, and yellow $F_\nu > 40$ Jy ms.}
    \label{fig:dectime}
\end{figure}

 We have focused our attention on the cumulative distributions in Fig.~\ref{fig:CDFs}, but note that our simulations naturally predict the detections as a function of time. As an example, to compare with Fig.~3 of
\citet{rn3}, in Fig.~\ref{fig:dectime} we show a set of randomly selected  repeaters from our reference model as a function of declination versus time. While we see a qualitative similarity to Fig.~3 of \citet{rn3}, we note that (as anticipated from Fig.~\ref{fig:CDFs}), our modeling approach results in an overabundance of sources at higher declinations than actually observed.
As seen by \citet{rn3}, it is notable that the higher declination sources benefit from a longer exposure time due to the unique optics of CHIME  resulting in more bursts observed. Further modeling, including different burst rate distributions would be a logical extension of this work.

One aspect we do consider in this paper, however, are some straightforward predictions for the number of repeaters we expect with future CHIME/FRB observations.
The power-law burst rate distribution we have considered from \citet{ClancyJames} produces a good match for the data through \citet{rn3}, which to date has reported a near-linear pattern of detection incidence. After scaling the number of sources to match the number of total number of cumulative detections by CHIME/FRB to date, our model predicts this almost linear trend will continue for the next decade,
in agreement with \citet{ClancyJames}. Any drop in this overall detection rate within the next few years would clearly challenge this model. 

\begin{figure*}
    \centering
    \includegraphics[width=0.9\textwidth]{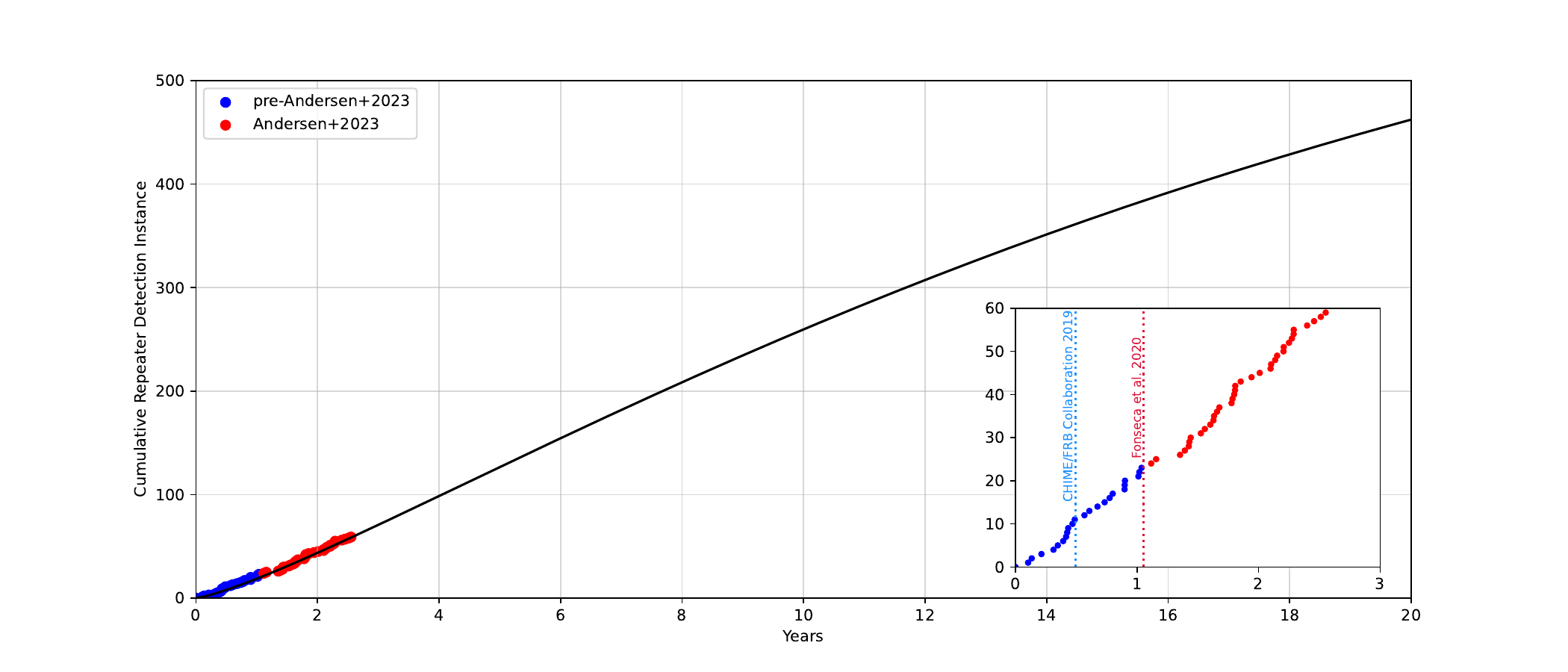}
    \caption{Cumulative repeater incidence for CHIME/FRB along with our model projection for the next two decades.  We scale the modeled detection incidence to the first 3.3 years of CHIME/FRB results. Inset: zoomed-in view of the current sample \citep{rn3} which shows an approximately linear trend.}
    \label{fig:CHIME_prediction}
\end{figure*}

\begin{figure*}
    \centering
    \includegraphics[width=0.9\textwidth]{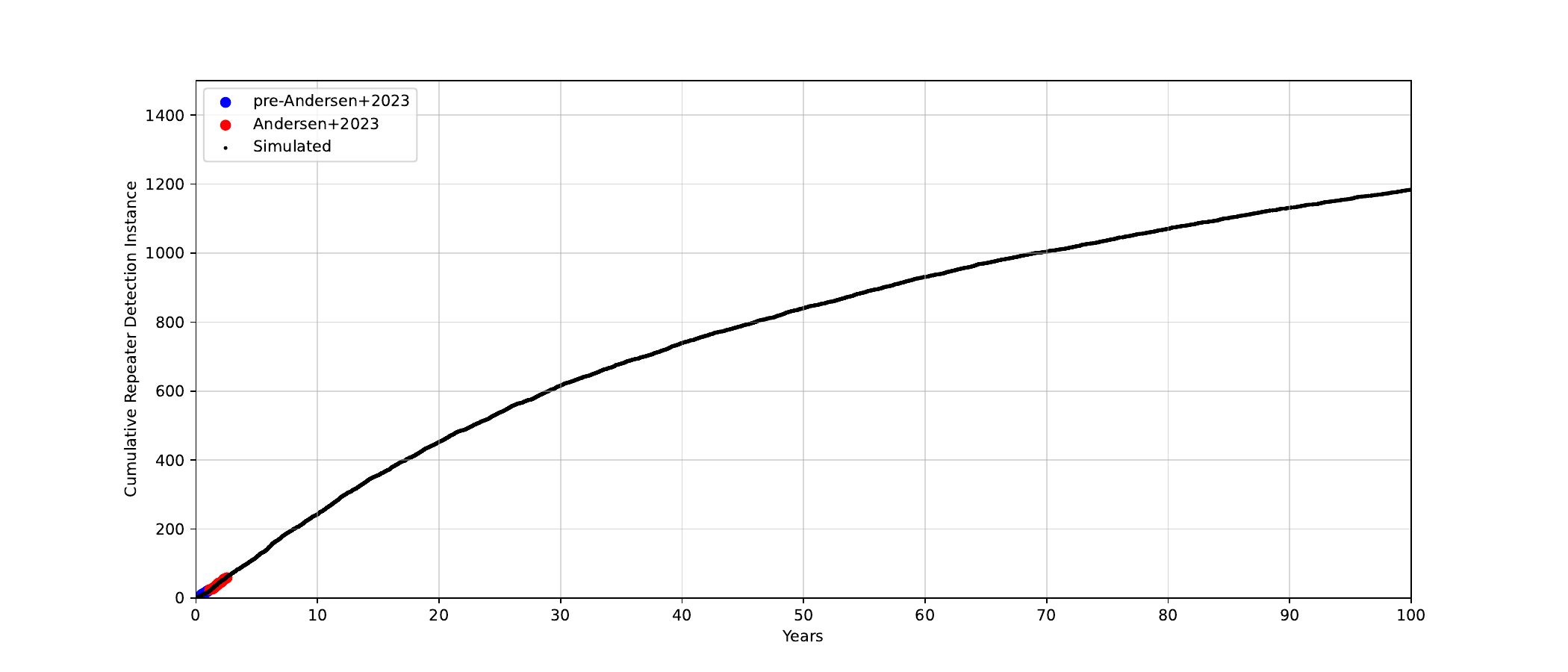}
    \caption{Model cumulative repeater incidence projection for CHIME/FRB extended out to a century of continued observing. We note the rapid decrease in detection rate between 20-40 years of elapsed observation, in contrast to the nearly linear trend projected for the first 20 years.}
    \label{fig:Extended_prediction}
\end{figure*}

In addition to the forward-modeled detection rates shown in Fig. \ref{fig:CHIME_prediction}, we extend our simulation to a full century of continued observations in Fig. \ref{fig:Extended_prediction}. While well beyond any conception for planned operations of CHIME/FRB, we wish to note the decrease in detections over very long cadences predicted by this model. Beyond the near-linear increase in new detections for the first 20 years, there is a notable decrease in detection rate between 20 and 40 years. While the detection rate trails off sufficiently such that the number of cumulative detections within the first 8 years equals that of the final 30 years of the simulation, no asymptotic total detectable repeaters value is identifiable on the considered timescale. This supports predictions of a consistently productive and growing repeater population detectable by CHIME/FRB across its operations.

\subsection{One-off FRBs in the observed population}

An interesting question that our modeling approach can address is what the number of apparent one-off sources in the current CHIME sample are actually repeaters. Since our modeling is exclusively assuming a repeating population, we automatically keep track of any simulated FRBs for which only one pulse is recorded over any given timescale. For the constant star formation case, we find that, over a three-year period, for every repeater observed, there are 2.8 apparently one-off sources. In other words, for the 60 repeaters currently observed by
\citet{rn3}, our model predicts that around 170 of the currently one-off sources in the CHIME/FRB catalog will repeat. Based on the projection shown in Fig.~\ref{fig:CHIME_prediction}, we anticipate that these sources are very likely to repeat within 5 yr. 

\subsection{Predictions for other instruments}

Our simulations can also be used to provide interesting predictions for hypothetical FRB search campaigns with two forthcoming next-generation radio observatories, the Deep Synoptic Array (DSA-2000)  and the Canadian Hydrogen Observatory and Radio-transient Detector (CHORD). These facilities, which are planned for construction and first light within the decade, are designed to have greater sensitivity than CHIME. DSA-2000 is projected to have a field of view FoV = 10.6 deg$^2$, with a fluence detection threshold for a 1 ms burst of 0.03~Jy~ms and a bandwidth of 850~MHz in the 1--2 GHz band \citep{DSA2000}. CHORD is slated to be the direct successor to CHIME, with FoV = 130 deg$^2$, a fluence detection threshold for a 1~ms burst of 0.1~Jy~ms, and a bandwidth of 1200~MHz in the 0.3--1.4~GHz band \citep{CHORD}. 

\begin{figure*}
    \centering
    \includegraphics[width=0.9\textwidth]{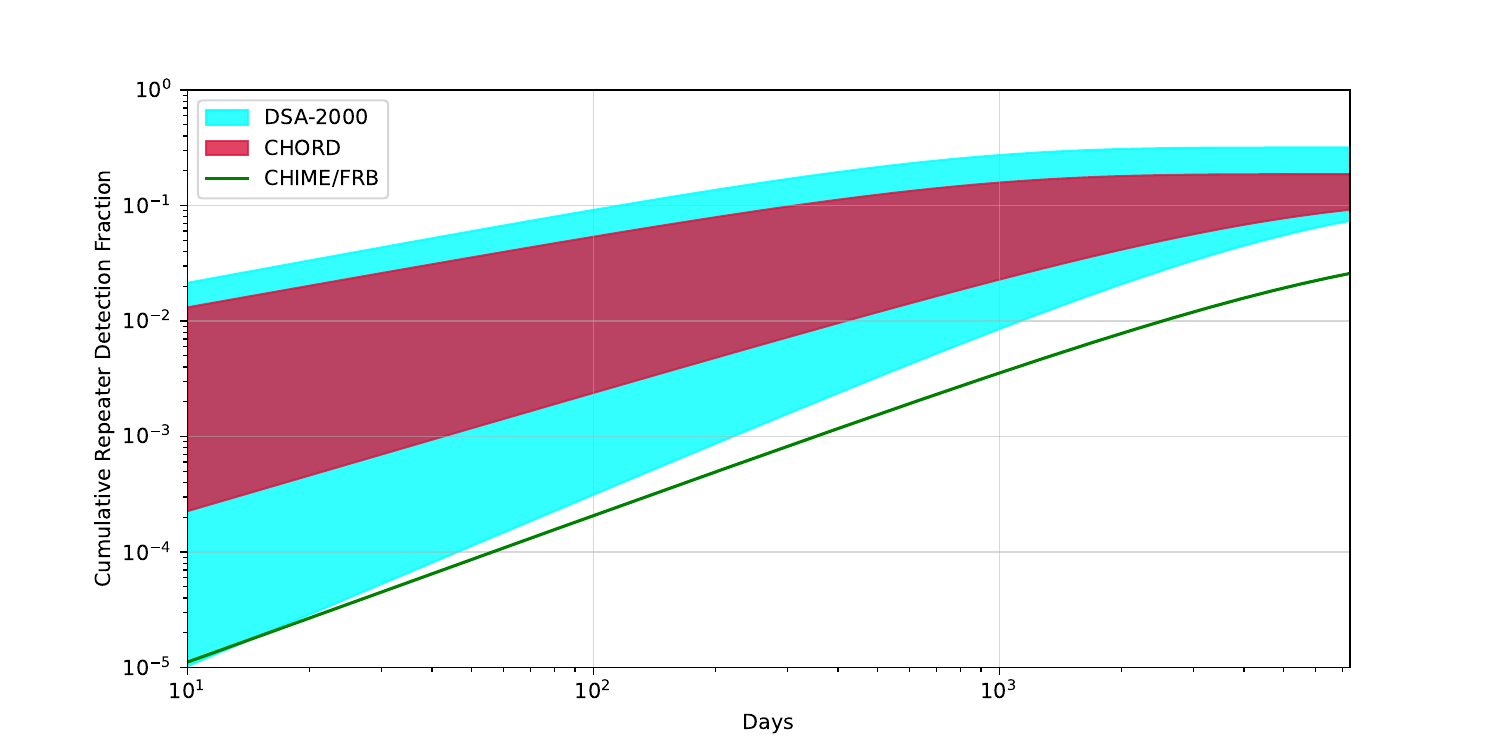}
    \caption{Cumulative detection predictions for DSA2000 and CHORD compared to CHIME/FRB prediction from Fig. \ref{fig:CHIME_prediction}. For display purposes, the model projections have been smoothed by fitting their cumulative incidences to gamma distributions. The upper and lower limits are chosen to reflect two extremes of observational strategies for observing a specific field of view.}
\label{fig:CombinedPredictionFits}
\end{figure*}

Since specifications for both DSA-2000 and CHORD are under development, we attempt to make a simple set of assumptions to illustrate the impacts of these sensitivity improvements. We proceed with two cases acting as upper and lower limits for the daily time in beam (exposure) for each source for each instrument, respectively. As an upper limit, we consider a sample of  sources constantly observed over the entire simulated campaign, corresponding to a circumpolar or never-set population always within the array's field of view, giving a daily exposure of 1 day. For a lower limit, we assume a population of sources at the celestial equator, such that their daily exposure is $\sqrt{\rm FoV}/360^{\circ}$~day. To enforce the completeness dropoff approaching the fluence limit, we scale the fit for $p(F_\nu)$ assuming an order of magnitude drop-off between the lower fluence limit and completeness limit ($F_{\rm lower} \sim 10 F_{\rm complete}$) for each instrument. This is likely to be a conservative estimate as these facilities will be more sensitive to bursts within $z = 0.5$ than CHIME. 

We show the range between these limiting cases for exposure in Fig.~\ref{fig:CombinedPredictionFits}. While we do not have a baseline for the size of population observed at higher sensitivities, as the scaling of the population discussed in section 5.2 is unique to a population observed with CHIME's sensitivity and unique declination-dependence not shared by these next-generation facilities, we can still make predictions for the detection rates by these observatories relative to the simulated population size.  Fig.~\ref{fig:CombinedPredictionFits} shows the cumulative repeater detections as a fraction of the total population. Unlike the case for CHIME/FRB, the more sensitive instruments appear to show a saturation in the number of repeating FRBs observed starting at around 1000~days. We emphasize here that our projections are limited to the FRB population within $z = 0.5$. The higher sensitivity of both DSA-2000 and CHORD appear to lead to a more complete census of the FRB population in this redshift range.

While the details of the putative surveys with DSA-2000 and CHORD are overly simplistic, and are restricted to the repeating FRB population with $z<0.5$, the results shown here highlight the potential for these and other future facilities with similar sensitivity to probe this population. Our results indicate that future facilities could potentially discriminate between population models due the stronger dependence on the number of detections with time.
Further simulations, particularly exploring the impacts of more realistic DSA-2000 and CHORD surveys on the population of FRBs with $z>0.5$ are encouraged.
In the meantime, we note that our predictions for CHIME are testable in the coming years and that CHIME's repeater sample is already a rich resource for studying the FRB population.

\section{Conclusions}
\label{conclusion}

We have described a simple simulation of the repeating FRB population that is based on the sample reported recently by \citet{rn3}. We find that this
sample of repeaters can be well described by simple model in which the sources are restricted to redshifts $z < 0.5$. Within this range, we cannot distinguish between a redshift distribution for the progenitors of FRB repeaters that follows the cosmic star formation, or is constant in comoving volume. As future CHIME/FRB observations are collected, as pointed out by \citet{ClancyJames}, we anticipate future studies being more sensitive to the population of FRBs with $z>0.5$.

In spite of the daily cadence of the CHIME/FRB observations, our models predict that the number of observed repeaters will not saturate significantly in the coming years and that the sample will grow in size approximately at the current rate of 20 new repeaters per year. A caveat to this prediction is its dependence on our assumptions. Future data releases from CHIME/FRB will allow us to constrain the model parameters/assumptions from this study. Among the future discoveries our models also predict in the next 5 years are around 170 sources that are currently in the CHIME/FRB sample as ``one-off'' FRBs.

Our simulation approach has tried to strike a balance between simplicity and rigor. We have attempted to incorporate the most important aspects of the population and detection process into our work, but have not tried to fine tune or do parameter estimation of the model parameters. Further modeling of the sample which explores the luminosity function, burst rate distribution and sky exposure are certainly warranted but beyond the scope of the current work. In particular, we have not accounted for the spectral behavior of the model FRBs, and the possibility of non-Poissonian burst distributions. Studies of this nature would be extremely valuable to better understand the growing population of repeating FRBs.

\acknowledgments

This research was carried out during a 10-week Research Experience for Undergraduates (REU) program at West Virginia University. We gratefully acknowledge the National Science Foundation's support of the REU under award number 1950617, as well as the NASA West Virginia Space Grant Consortium for their support. We thank Emmanuel Fonseca, Clancy James, Marcus Merryfield and Vicky Kaspi for useful discussions. We are also grateful to the referee for very useful comments on the manuscript.

%

\vspace{5mm}


\software{\texttt{astropy} \citep{2013A&A...558A..33A},  
          \texttt{ scipy} \citep{2020SciPy-NMeth}, 
          \texttt{ pygedm} \citep{pygedm}
          }

\bibliographystyle{mnras}
\bibliography{references.bib} 





\end{document}